\documentclass[12pt]{article}

\makeatletter
\@addtoreset{equation}{section}
\makeatother

\let\la=\label  
  
 \def\bd{\begin{document}} \def\ed{\end{document}}
\def\ds{\documentstyle} \let\fr=\frac \let\bl=\bigl \let\br=\bigr
\let\Br=\Bigr \let\Bl=\Bigl
\let\bm=\bibitem
\let\na=\nabla
\let\pa=\partial \let\ov=\overline
\newcommand{\be}{\begin{equation}}
\newcommand{\ee}{\end{equation}}
\def\ba{\begin{array}}
\def\ea{\end{array}}
\newcommand{\ho}[1]{$\, ^{#1}$}
\newcommand{\hoch}[1]{$\, ^{#1}$}
\newcommand{\bea}{\begin{eqnarray}}
\newcommand{\eea}{\end{eqnarray}}
\newcommand{\ra}{\rightarrow}
\newcommand{\lra}{\longrightarrow}
\newcommand{\Lra}{\Leftrightarrow}
\newcommand{\ap}{\alpha^\prime}
\newcommand{\bp}{\tilde \beta^\prime}
\newcommand{\tr}{{\rm tr} }
\newcommand{\Tr}{{\rm Tr} }
\newcommand{\NP}{Nucl. Phys. }
\newcommand{\tamphys}{\it
The Blackett Laboratory, Imperial College London,\\ Prince Consort Road, London SW7 2AZ}

\newcommand{\auth}{M. J. Duff\footnote{m.duff@imperial.ac.uk}}

\begin{document}
\hfill{}


\hfill{hep-th/0602160}

\vspace{24pt}

\begin{center}
{ \large {\bf Hidden symmetries of the Nambu-Goto action}}

\vspace{24pt}

\auth

\vspace{10pt}

{\tamphys}

\vspace{24pt}

\underline{ABSTRACT}

\end{center}

We organize the eight variables of the four-dimensional bosonic 
string $({\dot X}^{\mu}, X'^{\mu})$ into a $2 \times 2 \times 2$ 
hypermatrix $a_{AA'A''}$ and show that in signature $(2,2)$ the Nambu-Goto Lagrangian is 
given by $\sqrt{\rm{Det}~a}$ where Det is Cayley's hyperdeterminant. 
This is invariant not only under $[SL(2,R)]^{3}$ but also under 
interchange of the indices $A$, $A'$ and $A''$. This triality reveals hitherto hidden discrete 
symmetries of the Nambu-Goto action.

\vfill
\leftline{}

\newpage

\section{Cayley's hyperdeterminant}

In 1845 Cayley \cite{Cayley} generalized the determinant of a $2 \times 2$ matrix 
$a_{AA'}$
\[
{\rm det}~a=\frac{1}{2}\epsilon^{AB}\epsilon^{A'B'}a_{AA'}a_{BB'}
\]
\be
=a_{00}a_{11}-a_{01}a_{10}
\la{det}
\ee
to the {\it hyperdeterminant} of a $2 \times 2 \times 2$ {\it hypermatrix} 
$a_{AA'A''}$
\[
{\rm Det}~a=-\frac{1}{2}
\epsilon^{AB}\epsilon^{A'B'}\epsilon^{CD}\epsilon^{C'D'}\epsilon^{A''D''}\epsilon^{B''C''}
{a}_{AA'A''}{a}_{BB'B''}{a}_{CC'C''}{a}_{DD'D''}
\]
\[
=  a_{000}^2 a_{111}^2 + a_{001}^2 a_{110}^2 +
        a_{010}^2 a_{101}^2 + a_{100}^2 a_{011}^2 
\]
\[
        -2(a_{000}a_{001}a_{110}a_{111}+a_{000}a_{010}a_{101}a_{111}
\]
\[
        + a_{000}a_{100}a_{011}a_{111}+a_{001}a_{010}a_{101}a_{110}
\]
\[
        + a_{001}a_{100}a_{011}a_{110}+a_{010}a_{100}a_{011}a_{101}) 
\]
\be
        + 4 (a_{000}a_{011}a_{101}a_{110} + a_{001}a_{010}a_{100}a_{111})
\la{hyperdet}
\ee
The hyperdeterminant vanishes iff the following system of equations 
in six unknowns $p^{A},q^{A'},r^{A''}$ has a nontrivial solution, not 
allowing any of the pairs to be both zero:
\[
a_{AA'A''}p^{A}q^{A'}=0
\]
\[
a_{AA'A''}p^{A}r^{A''}=0
\]
\be
a_{AA'A''}q^{A'}r^{A''}=0
\ee
For our purposes, the important properties of the hyperdeterminant are that it is a quartic 
invariant under $[SL(2)]^{3}$ and under a triality that interchanges 
$A$, $A'$ and $A''$. 

We recently found a physical application of this hyperdeterminant \cite{Duff:2006uz} by 
associating the eight components of $a_{AA'A'}$ with the four electric and 
four magnetic charges of the STU black hole in four-dimensional string theory \cite{Duff:1995sm} 
and showing that its entropy \cite{Behrndt:1996hu} is given by
\be
 S= \pi \sqrt{-{\rm Det}~a}.
\ee  
The hyperdeterminant also makes it appearance in quantum information 
theory \cite{Miyake}.  Let the three qubit system $AA'A''$ be in a pure state $|\Psi\rangle$, 
and let the components of $|\Psi\rangle$ in the standard basis be 
$a_{AA'A''}$:
\begin{equation}
|\Psi\rangle = \sum_{AA'A''}a_{AA'A''}|AA'A''\rangle
\end{equation}
Then the three way entanglement of the three qubits $A$, $A'$ and 
$A''$ is given by 
the {\it 3-tangle} \cite{Coffman}
\be
\tau_{3}=4 |{\rm Det}~a|.
\ee
As far as we can tell \cite{Duff:2006uz}, the appearance of the Cayley 
hyperdeterminant in these two different contexts of stringy black hole entropy
(where the $a_{AA'A'}$ are integers and the symmetry is 
$[SL(2,Z)]^{3}$) and 3-qubit quantum entanglement (where the $a_{AA'A'}$ 
are complex numbers and the symmetry is 
$[SL(2,C]^{3}$) is a purely 
mathematical coincidence. Nevertheless, it has already provided 
fascinating new insights into the connections between strings, black holes, and 
quantum information \cite{Kallosh:2006zs}.

In this paper, we provide a third physical application by allowing 
$a_{AA'A''}$ to represent the eight variables of the 
four-dimensional bosonic string $({\dot X}^{\mu}, X'^{\mu})$ and 
showing that in signature $(2,2)$ the Nambu-Goto Lagrangian 
\cite{Nambu,Goto} is given by $\sqrt{\rm{Det}~a}$. Interchange of the indices $A$, $A'$ and 
$A''$ then reveals hitherto hidden discrete 
symmetries of the Nambu-Goto action.

\section{The Nambu-Goto action}
\la{action}

The Nambu-Goto action in a flat target space of signature $(2,2)$ is given by 
\be
I_{NG}= -\frac{T}{2}\int d\tau d\sigma \sqrt{- det~\gamma}
\la{nambu}
\ee
where
\be
\gamma_{A''B''}= \partial_{A''}X^{\mu}\partial_{B''}X^{\nu}\eta_{\mu\nu}
\la{gamma1}
\ee
and $\eta =diag~(1,1,-1,-1)$. We introduce a two-component notation
\be
{X}^{AA'}=\frac{1}{\sqrt{2}} 
\pmatrix{-{{X}}^{0}+{{X}}^{2} & {{X}}^{1}-{{X}}^{3} \cr 
         -{{X}}^{1}-{{X}}^{3} & -{{X}}^{0}-{{X}}^{2}}
\ee
and define
\be
a_{AA'A''}= \pmatrix{{\dot{X}}^{AA'}\cr X'^{AA'}},
\ee
where ${\dot X}=\partial X /\partial \tau$  and ${X'}=\partial X /\partial 
\sigma$. 
Hence
\be
\gamma_{A''B''}=\epsilon^{AB}\epsilon^{A'B'}a_{AA'A''}a_{BB'B''}
\ee
Then from (\ref{det}) and (\ref{hyperdet})
\be
{\rm det}~\gamma=-{\rm Det}~a
\ee
The $SL(2,R)$ acting on the index $A$ and the 
$SL(2,R)$ acting on the index $A'$ are just the $O(2,2) \sim SL(2,R) \times 
SL(2,R)$ spacetime symmetry, while the $SL(2,R)$ acting on the index 
$A''$ is the worldsheet symmetry
\be
\pmatrix{{\dot{X}}^{\mu}\cr X'^{\mu}}\rightarrow 
\pmatrix{  d & c \cr b & a}\pmatrix{{\dot{X}}^{\mu}\cr X'^{\mu}}
\ee
where $a,b,c,d$ are constants satisfying $ad-bc=1$. 

\newpage

\section{Discrete symmetries}
Interchanging $A$ 
and $A'$ is just reverses the sign of ${\dot X}^{1}$ and $X'^{1}$ but 
the 4 transformations 
\[
a_{AA'A''} \rightarrow a_{A''AA'}
\]
\[
a_{AA'A''} \rightarrow a_{A'A''A}
\]
\[
a_{AA'A''} \rightarrow a_{A''A'A}
\]
\be
a_{AA'A''} \rightarrow a_{AA''A'}
\ee
represent new discrete symmetries which are given explicitly by
\be
\left(
\begin{array}{c}
{\dot{X}}^{0}\\
{\dot{X}}^{1}\\
{\dot{X}}^{2}\\
{\dot{X}}^{3}\\
X'^{0}\\
X'^{1}\\
X'^{2}\\
X'^{3}
\end{array}
\right)
\rightarrow
\frac{1}{2}
\left(
\begin{array}{c}
{\dot{X}}^{0}-X'^{1}-{\dot{X}}^{2}+X'^{3}\\
-X'^{0}-{\dot{X}}^{1}+X'^{2}+{\dot{X}}^{3}\\
-{\dot{X}}^{0}-X'^{1}+{\dot{X}}^{2}+X'^{3}\\
X'^{0}-{\dot{X}}^{1}-X'^{2}+{\dot{X}}^{3}\\
X'^{0}+{\dot{X}}^{1}+X'^{2}+{\dot{X}}^{3}\\
{\dot{X}}^{0}-X'^{1}+{\dot{X}}^{2}-X'^{3}\\
X'^{0}-{\dot{X}}^{1}+X'^{2}-{\dot{X}}^{3}\\
{\dot{X}}^{0}+X'^{1}+{\dot{X}}^{2}+X'^{3}
\end{array}
\right)\ 
\ee

\be
\left(
\begin{array}{c}
{\dot{X}}^{0}\\
{\dot{X}}^{1}\\
{\dot{X}}^{2}\\
{\dot{X}}^{3}\\
X'^{0}\\
X'^{1}\\
X'^{2}\\
X'^{3}
\end{array}
\right)
\rightarrow
\frac{1}{2}
\left(
\begin{array}{c}
{\dot{X}}^{0}+X'^{1}-{\dot{X}}^{2}+X'^{3}\\
 X'^{0}-{\dot{X}}^{1}-X'^{2}-{\dot{X}}^{3}\\
-{\dot{X}}^{0}+X'^{1}+{\dot{X}}^{2}+X'^{3}\\
X'^{0}+{\dot{X}}^{1}-X'^{2}+{\dot{X}}^{3}\\
X'^{0}-{\dot{X}}^{1}+X'^{2}+{\dot{X}}^{3}\\
-{\dot{X}}^{0}-X'^{1}-{\dot{X}}^{2}+X'^{3}\\
X'^{0}+{\dot{X}}^{1}+X'^{2}-{\dot{X}}^{3}\\
{\dot{X}}^{0}-X'^{1}+{\dot{X}}^{2}+X'^{3}
\end{array}
\right)\ 
\ee

\be
\left(
\begin{array}{c}
{\dot{X}}^{0}\\
{\dot{X}}^{1}\\
{\dot{X}}^{2}\\
{\dot{X}}^{3}\\
X'^{0}\\
X'^{1}\\
X'^{2}\\
X'^{3}
\end{array}
\right)
\rightarrow
\frac{1}{2}
\left(
\begin{array}{c}
{\dot{X}}^{0}-X'^{1}-{\dot{X}}^{2}+X'^{3}\\
X'^{0}+{\dot{X}}^{1}-X'^{2}-{\dot{X}}^{3}\\
-{\dot{X}}^{0}-X'^{1}+{\dot{X}}^{2}+X'^{3}\\
X'^{0}-{\dot{X}}^{1}-X'^{2}+{\dot{X}}^{3}\\
X'^{0}+{\dot{X}}^{1}+X'^{2}+{\dot{X}}^{3}\\
-{\dot{X}}^{0}+X'^{1}-{\dot{X}}^{2}+X'^{3}\\
X'^{0}-{\dot{X}}^{1}+X'^{2}-{\dot{X}}^{3}\\
{\dot{X}}^{0}+X'^{1}+{\dot{X}}^{2}+X'^{3}
\end{array}
\right)\ 
\ee

\be
\left(
\begin{array}{c}
{\dot{X}}^{0}\\
{\dot{X}}^{1}\\
{\dot{X}}^{2}\\
{\dot{X}}^{3}\\
X'^{0}\\
X'^{1}\\
X'^{2}\\
X'^{3}
\end{array}
\right)
\rightarrow
\frac{1}{2}
\left(
\begin{array}{c}
{\dot{X}}^{0}+X'^{1}-{\dot{X}}^{2}+X'^{3}\\
- X'^{0}+{\dot{X}}^{1}+X'^{2}+{\dot{X}}^{3}\\
-{\dot{X}}^{0}+X'^{1}+{\dot{X}}^{2}+X'^{3}\\
X'^{0}+{\dot{X}}^{1}-X'^{2}+{\dot{X}}^{3}\\
X'^{0}-{\dot{X}}^{1}+X'^{2}+{\dot{X}}^{3}\\
{\dot{X}}^{0}+X'^{1}+{\dot{X}}^{2}-X'^{3}\\
X'^{0}+{\dot{X}}^{1}+X'^{2}-{\dot{X}}^{3}\\
{\dot{X}}^{0}-X'^{1}+{\dot{X}}^{2}+X'^{3}
\end{array}
\right)\ 
\ee

\section{Conclusions}

One way to understand this triality is to think of the string as 
having three different worldsheets (Alice, Bob and Charlie) with three 
different metrics
\[
\alpha_{AB}=\epsilon^{A'B'}\epsilon^{A''B''}a_{AA'A''}a_{BB'B''}
\]
\[
\beta_{A'B'}=\epsilon^{A''B''}\epsilon^{AB}a_{AA'A''}a_{BB'B''}
\]
\be
\gamma_{A''B''}=\epsilon^{AB}\epsilon^{A'B'}a_{AA'A''}a_{BB'B''}
\ee
All are equivalent, however, since
\be
det~\alpha=det~\beta=det~\gamma=-{\rm Det}~a
\ee
It is remarkable that the Nambu-Goto action, first written down in 
1970 can still reveal some secrets, although the $(2,2)$ signature means that their 
physical significance is not obvious. It is worth remarking, however, that this 
does form the bosonic sector of the N=2 critical superstring. See 
\cite{Ooguri}, for example.

Finally we recall \cite{Kallosh:2006zs} that the $E_{7(7)}$ Cartan 
invariant $J_{4}$, depending on $28 +28$ variables, reduces to Cayley's 
hyperdeterminant in a canonical 
basis. So one could generalize the Nambu-Goto action to 
\be
I_{NG}= -\frac{T}{2}\int d\tau d\sigma \sqrt{J_{4}}
\la{nambu2}
\ee
Bearing in mind the embeddings
\be
E_{7(7)} \supset SO(6,6) \times SL(2,R) \supset SO(2,2) \times SL(2,R)
\ee
this would describe a string in 28 dimensions with reduced Lorentz group 
$SO(6,6)$, rather than $SO(2,2)$.  It may seem strange to embed the 
Lorentz group in a bigger group, but this was already true for the 
conventional Nambu-Goto theory (\ref{nambu}), since $[SL(2,R)]^{3}$ 
and triality form a semi-direct product.

Perhaps our familiarity with string theory will allow us to gain yet 
more insights into the other two contexts in which Cayley's 
hyperdeterminant makes it appearance, namely black holes and quantum 
information, and vice-versa. 

\section{Acknowledgements}

I am grateful to Dimitrios Giataganas for pointing out some 
typographical errors in the original version. I also thank Jussi 
Kalkkinen and Tom Kibble for the observation that $[SL(2,R)]^{3}$ 
and triality do not commute.
 


\begin{thebibliography}{10}

\bibitem{Cayley}
A. Cayley,
``On the theory of linear transformations,''
Camb. Math. J. 4 193-209,1845.

	
\bibitem{Duff:2006uz}
 M.~J.~Duff,
``String triality, black hole entropy and Cayley's hyperdeterminant,''
 arXiv:hep-th/0601134.
  
\bibitem{Duff:1995sm}
 M.~J.~Duff, J.~T.~Liu and J.~Rahmfeld,
``Four-dimensional string-string-string triality,''
  Nucl.\ Phys.\ B {\bf 459}, 125 (1996)
  [arXiv:hep-th/9508094].
  
\bibitem{Behrndt:1996hu}
  K.~Behrndt, R.~Kallosh, J.~Rahmfeld, M.~Shmakova and W.~K.~Wong,
  ``STU black holes and string triality,''
  Phys.\ Rev.\ D {\bf 54}, 6293 (1996)
  [arXiv:hep-th/9608059].

\bibitem{Miyake}
A. Miyake and M. Wadati,
``Multipartite entanglement and hyperdeterminants,''
[arXiv:quant-ph/02121146].

\bibitem{Coffman}
V. Coffman, J. Kundu and W. Wooters,
``Distributed entanglement,''
Phys. Rev. A61 (2000) 52306,
[arXiv:quant-ph/9907047].

\bibitem{Kallosh:2006zs}
  R.~Kallosh and A.~Linde,
  ``Strings, black holes, and quantum information,''
  arXiv:hep-th/0602061.

\bibitem{Nambu}
Y. Nambu,
``Duality and hydrodynamics,''
Lectures at the Copenhagen conference, 1970.

\bibitem{Goto}
T. Goto,
``Relativistic quantum mechanics of one-dimensional mechanical
continuum and subsidiary condition of dual resonance,''
Prog.  Theor.  Phys.  {\bf 46} (1971) 1560.

\bibitem{Ooguri}
H. Ooguri and C. Vafa
``Geometry of $N=2$ strings,''
Nucl. Phys. B361 (1991) 469.

\end{thebibliography}
\end{document}